\documentclass[12pt]{article}
\def\lsim{\mathrel{\rlap{
\lower4pt\hbox{\hskip-3pt$\sim$}}
    \raise1pt\hbox{$<$}}}     
\def\gsim{\mathrel{\rlap{
\lower4pt\hbox{\hskip-3pt$\sim$}}
    \raise1pt\hbox{$>$}}}     
\usepackage{graphicx}

\begin{document}
\input{epsf}

\title{ Strangeness Production  in Nuclear Matter and Expansion Dynamics }
\author {
V.D.~Toneev$^{a,b}$, E.G.~Nikonov$^{a,b }$, \\
 B.~Friman$^{ a}$,
 W.~N\"orenberg$^{a}$,
  K.~Redlich$^{c,d}$\\
$^{a} $
 Gesellschaft f\"ur Schwerionenforschung GSI,\\
D--64291 Darmstadt, Germany \\
 $^{b}$  Joint Institute for Nuclear Research,
  141980 Dubna, \\  Moscow Region, Russia \\
$^{c}$ Fakult\"at f\"ur Physik, Universit\"at Bielefeld,\\
  D--33501 Bielefeld, Germany\\
$^{d}$ Institute of Theoretical Physics, University of Wroc\l aw,\\
PL--50204
Wroc\l aw, Poland \\
 } \maketitle

\begin{abstract}
Thermodynamical properties of hot and dense nuclear matter are
analyzed and compared for different equations of state (EoS). It
is argued that the softest point of the equation of state  and the
strangeness separation on the phase boundary can manifest
themselves in observables. The influence of the EoS and the order
of the phase transition  on the expansion dynamics of nuclear
matter and strangeness excitation function is analyzed. It is
shown that  bulk properties of strangeness production in A--A
collisions depend only   weakly on the particular form of the EoS.
The predictions of   different models are related with
experimental data on strangeness production.
\end{abstract}


\addvspace{5mm} \newpage

\section{Introduction}

The quest for the deconfinement transition, the phase transition
from a confined hadronic phase to a deconfined quark--gluon phase
(the so called quark--gluon plasma, QGP), remains a major
challenge in strong interaction physics \cite{HJ00}. Over the past
two decades a lot of effort has gone into the exploration of this
transition and its possible manifestations in relativistic heavy
ion collisions, in neutron stars as well as in the early universe.
Relativistic heavy ion collisions offers a unique opportunity to
reach states with temperatures and energy densities exceeding the
critical values, $T_c\sim 170 \ MeV$ and $\varepsilon_c \sim 0.6 \
GeV/\mbox{fm}^3$, specific for the deconfinement phase transition
\cite{crit}. Thus, it is likely that color degrees of freedom play
an important role already at SPS and RHIC energies \cite{hadro}.
Various signals for the formation of a quark--gluon plasma in such
collisions have been discussed and probed in
experiments~\cite{HJ00,RHIC}.

Enhanced production of strangeness relative to proton--proton and
proton--nucleus collisions was one of the conjecture signals   of
the quark--gluon plasma formation  in heavy ion collisions
\cite{KMR86}. The original idea behind the strangeness enhancement
is that strange and antistrange quarks are easily created in a
quark--gluon plasma, while in the hadronic phase strangeness
production is suppressed. The dominant reaction in the plasma is
$gg\to s\bar{s}$. Furthermore, since the strange quark mass is not
larger than $T_c$, one expects the strange degrees of freedom to
equilibrate in the quark--gluon plasma. Although, a heavy ion
collision at high energies is a highly non--equilibrium process,
the hadron yields (including strange particles)  measured in the
energy range from SIS to RHIC~\cite{heppe,COR99,rev,Red01,metag}
are remarkably well described in the thermal model assuming
chemical equilibrium at freeze out. This indicates that collective
effects play an important role in the production of strangeness.
On the other hand, elaborate microscopic transport models do not
provide a quantitative explanation of the excitation functions for
strange particles in this energy range. In the hadron string
dynamics model~\cite{Cas99} one finds a too small $K^+/\pi^+$
ratio around AGS energies, while in RQMD~\cite{DO00} the yield is
overestimated at SIS and too small at SPS energies.

The aim of this paper is to explore global effects of strangeness
production in hot and dense  nuclear matter within a collective
approach. Our starting point is an equation of state (EoS) with a
deconfinement phase transition. Since strangeness is conserved at
the time scales relevant for heavy ion collisions, a strangeness
chemical potential is introduced. We examine various
phenomenological models for the equation of state, which differ in
the order of the deconfinement phase transition: a first order
transition (the two--phase bag model), a crossover--type
transition (the statistical mixed--phase model) and no phase
transition (pure hadronic models). The consequences of strangeness
separation and softening of the equation of state are discussed.
Furthermore, the manifestation of the order of the deconfinement
phase transition in the expansion dynamics and the bulk
strangeness production is studied. The predictions obtained with
different equations of state are related with experimental
excitation functions for relative strange particle abundances.

\section{Modelling the equation of state of strongly interacting matter }

The EoS of strongly interacting matter can in general be obtained
by first principal calculations within lattice gauge theory
\cite{lat}. The thermodynamics and the order of the phase
transition in QCD is rather well established for two and three
light quark flavour in lattice calculations. However, the
physically relevant situation of two light (u,d) and a heavy (s)
quark is still not well described  within lattice approach. In
particular, the existence of a phase transition and its order in
2+1 flavour QCD is not yet known. In addition most of the lattice
calculations are performed for vanishing net baryon number
density. Only recently, first results on the EoS with non zero
baryon chemical potential have been obtained  on the lattice
\cite{fodor}. However, these studies have so far been performed
with large   quark masses which distort the physical EoS. Thus,
lattice results can
 still not  be used directly in  physical  applications.

Lacking  lattice QCD results for the EoS at finite baryon density
$n_B$ with  physically relevant values of the quark masses,  a
common approach is to construct a phenomenological equation of
state for strongly interacting matter. This EoS should be
constrained by  existing lattice results and should also reproduce
the two--phase structure of QCD. Here we
  construct different models for  QCD thermodynamics and
 study  their physical implications with particular emphasis on
 strangeness production and evolution in heavy ion collisions.

A recent analysis of the lattice EoS  \cite{our1,our2},  shows
that in the low temperature phase, hadrons and resonances are the
relevant degrees of freedom. The hadron resonance gas, with a
modified mass spectrum to account for the unphysical values of the
quark masses used in the lattice calculations, was shown to
reproduce the bulk thermodynamic properties of QCD,  obtained on
the lattice with different numbers of quark flavors as well as at
finite  and vanishing net   baryon density \cite{our1,our2}.

Lattice calculations show that, at very large temperature the
thermodynamical observables approach the Stefan--Boltzmann limit
of an ideal gas of quarks and gluons, both at finite as well as
vanishing net baryon density. The remaining $\sim 20\%$
discrepancy at $T>2T_C$ is understood by systematic contributions
in self--consistent implementation of quasiparticle masses in the
HTL--reassumed   perturbative    QCD \cite{jp}. To describe the
thermodynamics near the phase transition additional model
assumptions are necessary \cite{quasi,pis}.

From the above discussion it is clear that the most
straightforward model for the EoS would be a non-interacting
hadron resonance gas in the low temperature phase and ideal quark
gluon--plasma in the color deconfined phase. These phases are
matched at the phase transition boundary by means of the Gibbs
phase equilibrium conditions. By construction, this approach
yields a first order phase transition. Such an EoS  with strange
degrees of freedom has  frequently been used in the literature
\cite{LRBH86,LZB87,GKS87,BFKS89,CSSW92,CGSS93} and is also a
standard input in   hydrodynamic simulations of heavy ion
collisions \cite{HS95,BD00}. However, in order to obtain a
reasonable phase diagram  one has  to include   short--range
repulsive interactions between hadronic constituents. In general
this can be realized by introducing short--range repulsion  in a
thermodynamically consistent approach
\cite{NST98,TNS98,GY95,BST01}.

We note that  according to Gibbs phase rule~\cite{LL}, the number
of thermodynamic degrees of freedom that may be varied without
destroying the equilibrium of a mixture of $r$ phases, with $n_c$
conserved charges is ${ \cal N}=n_c+2-r$. For the hadron--quark
deconfinement transition under consideration $r=2$. If the baryon
number is the only conserved quantity, $n_c=1$ and ${\cal N}=1$.
Thus, the phase boundary is one--dimensional, i.e. a line. The
Maxwell construction for a first order phase transition
corresponds just to this case  $r=2$ and $n_c=1$. When both the
baryon number and strangeness are conserved ($n_c=2$), one has
${\cal N}=2$ and therefore the phase boundary is in general  a
surface. In such a system, a standard Maxwell construction is not
possible~\cite{Glend92}.

To account for the uncertainties in the order of the phase
transition in 2+1 flavour QCD and also  for the deviation of the
equation of state from an  ideal gas  near the deconfinement
transition
 we employ  the EoS  of the mixed phase model
\cite{NST98,TNS98}.  In this model it is assumed that unbound
quarks and gluons { may coexist} with hadrons forming a {
homogeneous} mixture. This model is thermodynamically consistent
and reproduces the lattice EoS obtained  in the pure gauge theory
as well as in two flavour QCD. Furthermore,  the order of the
phase transitions in the mixed phase model  depends on the
strength of the interaction between the phases. In this approach
we can  study the importance of the order of the phase transition
on strangeness production and on the evolution of heavy ion
collisions.

In the following  we discuss first the basic thermodynamical
properties of these different models of the EoS and indicate
relevant differences in their predictions.

\subsection{Two--phase bag model}
In the two--phase (2P) model~\cite{Cleym},  the deconfinement
phase transition is determined  by matching the EoS of a
relativistic gas of hadrons and resonances, with repulsive
interactions at short distances, to that of an ideal gas of quarks
and gluons. The change in vacuum energy in the plasma phase is
parameterized by a bag constant $B$. We work in the grand
canonical ensemble and account for all hadrons with mass $m_j <
1.6 \mbox ~{GeV}$, including the strange particles and resonances
with strangeness $s_j=\pm 1,\pm 2,\pm 3$. The density of particle
species $j$ is then
\begin{eqnarray}
n_j(T,\mu_j )\equiv n_j(T,\mu_B,\mu_S )&=& v \
n_j^{id}(T,\mu_B,\mu_S) \nonumber \\ &=& \frac{v \
g_j}{2\pi^2}\int_0^{\infty} dk\ k^2 \ f_j(k,T,\mu_B,\mu_S)~,
\label{eqt1}
\end{eqnarray}
where
\begin{eqnarray}
 f_j(k,T,\mu_B,\mu_S) = \left[ \  exp \left( \frac{\sqrt{k^2 + m_j^2}- b_j\mu_B
-s_j\mu_S }{T} \right)
 \pm 1 \right]^{-1}
\label{eqt2}
\end{eqnarray}
is the momentum distribution function for fermions ($+$) and
bosons ($-$) while $g_j$ is the spin--isospin degeneracy factor.
The chemical potential $\mu_j$   is related to the baryon
($\mu_B$) and strangeness ($\mu_S$) chemical potentials
\begin{equation}
\mu_j = b_j \ \mu_B+s_j \ \mu_S\,, \label{eqt1a}
\end{equation}
where $b_j$ and $s_j$ are the baryon number and strangeness of the
particle. The quantity $ n_j^{id}$ corresponds to the number
density of an ideal point--like hadron gas (IdHG). The factor
\begin{equation}
v \equiv v(T,\mu_B,\mu_S )=1/[1+\sum_j \ v_{0j} \
n_{j}^{id}(T,\mu_B,\mu_S )] \label{eqt3}
\end{equation}
reduces the volume available   for hadrons due to their short
range repulsion determined  by the eigenvolume $v_{0j}=(1/2)(4\pi
/3)(2r_{0j})^3$ ~\cite{LL}. We choose the effective interaction
radius $r_{0j}\sim 0.5 \mbox{fm}$ for all hadrons. Following
(\ref{eqt1}), the baryon density  and strangeness in the hadronic
phase can be expressed as
\begin{eqnarray}
\label{eqt1b}
n_B^H &=&  \sum_{j \in h} b_j \  n_j(T,\mu_B,\mu_S )~, \\
n_S^H &=&  \sum_{j \in h} s_j \  n_j(T,\mu_B,\mu_S ) \label{eqt1c}
\end{eqnarray}
where the sum is taken over all hadrons and resonances. Similarly,
the energy density of species $j$ is given by
\begin{eqnarray}
\varepsilon_j(T,\mu_B,\mu_S )&=& v \
\varepsilon_j^{id}(T,\mu_B,\mu_S)
 \nonumber \\ &=&
\frac{v \ g_j}{2\pi^2}\int_0^{\infty} dk\ k^2  \sqrt{k^2+m_j^2} \
f_j(k,T,\mu_B,\mu_S)~. \label{eqt3a}
\end{eqnarray}

In early studies~ \cite{Cleym,excl_v}, the excluded volume
correction $v$ was implemented in the same way for all
thermodynamic quantities of the hadron gas, including the pressure
\begin{equation}
 p^H(T,\mu_B,\mu_S) = \sum_{j \in h} p_j(T,\mu_B,\mu_S)\,,
\label{eqt4}
\end{equation}
where the partial pressures are given by
\begin{eqnarray}
p_j(T,\mu_B,\mu_S )&=& v \ p_j^{id}(T,\mu_B,\mu_S) \nonumber \\
&=& \frac{v \ g_j}{6\pi^2}\int_0^{\infty} dk\
\frac{k^4}{\sqrt{k^2+m_j^2}} \ f_j(k,T,\mu_B,\mu_S)~. \label{eqt5}
\end{eqnarray}
However, this expansion for the  pressure is not thermodynamically
consistent with the    charge   (\ref{eqt1b}-\ref{eqt1c}) as well
as  the energy density (\ref{eqt3a}). In  Ref.~\cite{RGSG91}  it
was shown  that,  it is possible to account for a
thermodynamically consistent implementation of the excluded volume
corrections. In this approach the pressure is given by that of an
ideal gas with modified chemical potentials
\begin{equation}
 p^H(T,\mu_B,\mu_S) =  \sum_{j \in h} p_j^{id}(T,\tilde{\mu_j})
\label{eqt6}
\end{equation}
where
\begin{equation}
 \tilde{\mu_j} = b_j \ \mu_B +s_j \ \mu_S -v_{0j} \ p^H(T,\mu_B,\mu_S)~.
\label{eqt7}
\end{equation}
The remaining  thermodynamic quantities are obtained with the
excluded volume correction given above by taking  the
corresponding derivatives of the pressure. Thus, in this approach
all fundamental thermodynamic relations are
fulfilled~\cite{RGSG91}. We shall refer to
Eqs.(\ref{eqt4},\ref{eqt5}) and Eqs.(\ref{eqt6},\ref{eqt7}) as
two--phase thermodynamically { inconsistent} (2PIN) and {
consistent} (2PC) model, respectively. Note that such an equation
of state may violate causality at high densities, because an
extended rigid body is incompatible with the basic principles of
relativity.

The QGP phase is described as a gas of non--interacting point-like
quarks, antiquarks and gluons.  The non--perturbative effects
associated with confinement  are described by  the constant vacuum
energy $B$. The pressure in the plasma  is then given by
\begin{equation}
 p^Q(T,\mu_B,\mu_S) = p_g(T)+ \sum_{j \in q} p_j^{id}(T,\mu_B,\mu_S) - B~,
\label{eqt8}
\end{equation}
where the gluon
\begin{equation}
 p_g(T) = \frac{g_g \pi^2}{90} T^4  \hspace{10mm} (g_g=16)
\label{eqt9}
\end{equation}
and the quark pressure is obtained from  Eq.(\ref{eqt5}) for
 $u,d,s$ quarks and antiquarks. We use
the quark masses $m_u=m_d=5\mbox{MeV}$ and $m_s=150 \mbox{MeV}$
and the bag constant $B=(235\mbox{MeV})^4$  which  yields a
transition temperature $T_c \approx 160\mbox{MeV}$ in agreement
 with lattice calculations at $n_B=0$~ \cite{lat}. The energy
density of the plasma phase
\begin{equation}
 \varepsilon^Q(T,\mu_B,\mu_S) = \varepsilon_g(T)+
\sum_{j \in q} \varepsilon_j^{id}(T,\mu_B,\mu_S) + B~,
\label{eqt10}
\end{equation}
where the gluon contribution is given by
\begin{equation}
 \varepsilon_g(T) = 3 \ p_g(T) = \frac{g_g \pi^2}{30} T^4
\label{eqt11}
\end{equation}
and that of quark species $j$ is obtained from  Eq.(\ref{eqt3a})
with $v=1$. Analogously to Eqs.(\ref{eqt1b}) and (\ref{eqt1c}) the
densities of the conserved charges in the QGP phase are~:
\begin{eqnarray}
\label{eqt11b}
n_B^Q &=&  \sum_{j \in q} b_j \  n_j^{id}(T,\mu_B,\mu_S )~, \\
n_S^Q &=&  \sum_{j \in q} s_j \  n_j^{id}(T,\mu_B,\mu_S )~.
\label{eqt11c}
\end{eqnarray}
The  equilibrium between the plasma and the hadronic phase is
determined by the Gibbs conditions for thermal ($T^Q=T^H$),
mechanical ($p^Q=p^H$) and chemical ($\mu_B^Q =\mu_B^H, \ \mu_S^Q
=\mu_S^H$) equilibrium.
 At a given temperature $T$ and baryon chemical
potential $\mu_B$ the strange chemical potential $\mu_S$ is
obtained  by requiring that the net strangeness of the total
system vanishes. Thus, for  the total baryon density $n_B$ the
phase equilibrium requires that:
\begin{eqnarray}
\label{eqt12a}
p^H(T,\mu_B,\mu_S) &=& p^Q(T,\mu_B,\mu_S)~, \\
\label{eqt12b} n_B &=& \alpha \ n_B^Q(T,\mu_B,\mu_S ) + (1- \alpha
) \
n_B^H(T,\mu_B,\mu_S )~, \\
0 &=& \alpha \  n_S^Q(T,\mu_B,\mu_S ) + (1-\alpha ) \
n_S^H(T,\mu_B,\mu_S )~, \label{eqt12c}
\end{eqnarray}
where $\alpha = V_Q / V$ is the fraction of the volume occupied by
the plasma phase. The boundaries of the coexistence region are
found by putting $\alpha = 0$ (the hadron phase boundary) and
$\alpha = 1$ (the plasma boundary).

As mentioned above, the Maxwell construction is not appropriate in
a system where both baryon number and strangeness are conserved.
To illustrate this, we first analyze an approximate form of the
equation (\ref{eqt11c}) for strangeness conservation.
 We retain only the main terms and drop
those with $|s_j|> 1$:
\begin{equation}
\alpha \ (n_s-n_{\bar{s}}) = (1-\alpha) \
(n_{K}+n_{\bar{\Lambda}}+n_{\bar{\Sigma}}-
n_{\bar{K}}-n_{\Lambda}-n_{\Sigma})\,. \label{eqt13}
\end{equation}
In the Boltzmann approximation   the densities may be computed
analytically
\begin{equation}
n^{id}_j \approx  n^{B}_j= g_j \ (\frac{T^3}{2\pi^2})\  (
\frac{m_j}{T})^2 \ K_2(\frac{m_j}{T})\  \exp (\frac{\mu_j}{T})
\equiv g_j \ (\frac{T^3}{2\pi^2}) \ W_j \  \exp
(\frac{\mu_j}{T})~, \label{eqtB}
\end{equation}
and the strangeness chemical potential is obtained
as~\cite{RGSG91}:
\begin{eqnarray}
\mu_S = \frac{T}{2}\ln \frac{3 \alpha  W_s +v (1-\alpha)\left( W_K
\ e^{-\frac{\mu_B}{3T}}+(W_{\Lambda}
+3W_{\Sigma})e^{\frac{2\mu_B}{3T}}\right)} {3 \alpha W_s +v
(1-\alpha)\left( W_K \ e^{\frac{\mu_B}{3T}}+(W_{\Lambda}
+3W_{\Sigma})e^{-\frac{2\mu_B}{3T}}\right)} +\frac{\mu_B}{3}~.
\label{eqt14}
\end{eqnarray}
It is seen that at the {\it plasma boundary} ($\alpha=1$)
$\mu_S=\mu_B /3$ while $\mu_S \ne\mu_B /3$ at the {\it  hadron
boundary} ($\alpha=0$). This implies that not only $\mu_S$ but
also $\mu_B$ and the pressure changes  along isotherms in the
coexistence region. Hence, the standard Maxwell construction,
which interpolates the densities linearly between the pure phases,
is not adequate. The equations for phase equilibrium
(\ref{eqt12a}--\ref{eqt12c}) must be solved to obtain $\mu_S$ and
$\mu_B$ at every point in the coexistence region.

When two phases coexist, the system is in general not homogeneous
as the phases    occupy a separate domains in space. We do not
explicitly account for such domains structure nor  for a possible
surface energy contribution to the equation of state. The only
consequence of the phase separation in this calculations is that
the interactions between particles in the plasma and hadronic
phase are excluded. This is different in the statistical mixed
phase model discussed in the next section.

\begin{figure}[htb]
\begin{center}
\includegraphics[height=10cm, angle=-90,clip]{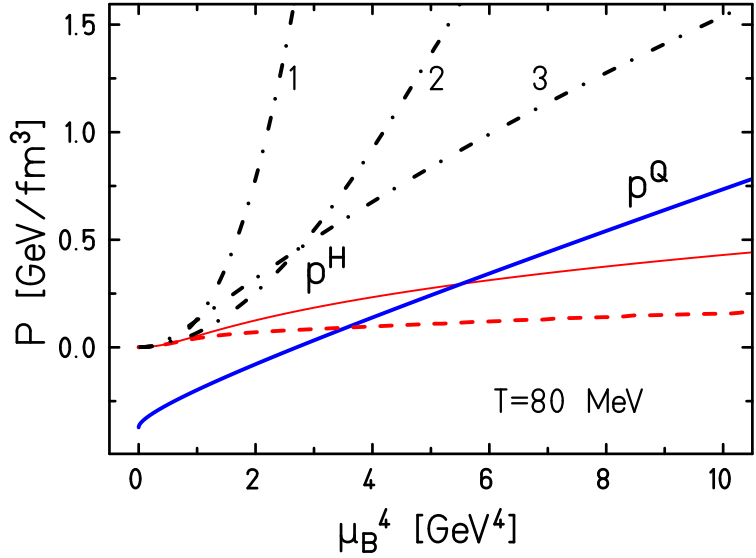}
\caption[C1]{Pressure { versus} baryon chemical potential for
fixed $T=80 \ MeV$ and for $\mu_S=\mu_B/3$. The  thin--line is the
hadronic and the thick--line the quark phase in 2PC model. The
dashed--dotted (1) line and dashed--line are an ideal gas model
results without and with repulsion in the 2PIN model,
respectively. The line (2) is obtained as line (1) but with fewer
hadronic resonances. The line (3) is calculated within a
mean--field approximation of the
 Zimanyi model~\cite{Zim} (see text).}
\label{fig1}
\end{center}
\end{figure}

The solution of the Gibbs conditions (\ref{eqt12a}--\ref{eqt12c})
is shown  in Fig.~\ref{fig1} for the plasma and hadron phase
pressure  versus $\mu_B^4$ at fixed $T=80 \mbox{MeV}$ and
$\mu_S=\mu_B/3$. The crossing  of the quark and hadronic pressure
corresponds to the transition point at the plasma boundary. In
this special case the condition $\mu_S=\mu_B/3$ guarantees
strangeness neutrality. In general, however,   for $\alpha \neq
1$, $\mu_S$ must be chosen such that the strangeness of the total
system of quarks and hadrons vanishes. This requires an iterative
solution of the equations (\ref{eqt12a}--\ref{eqt12c}). Away from
the transition point, the system is in the phase with higher
pressure $p$ (lower free energy). Fig.~\ref{fig1} also shows that
there is no deconfinement transition if the hadronic phase is
described as a gas of point--like particles~\cite{Cleym}. The
situation is not improved by including more resonances. On the
contrary, the larger the set of hadronic resonances is, the higher
is the pressure at a given baryon chemical potential. However, the
inclusion of repulsive interactions between hadrons leads to a
reduction of the hadron pressure $p^H$ at fixed baryon chemical
potential. Consequently, a short--range repulsion between hadrons
stabilizes the quark--gluon plasma at high densities.

\begin{figure}[htbp]
\begin{center}
\includegraphics[height=11.cm, angle=-90,clip]{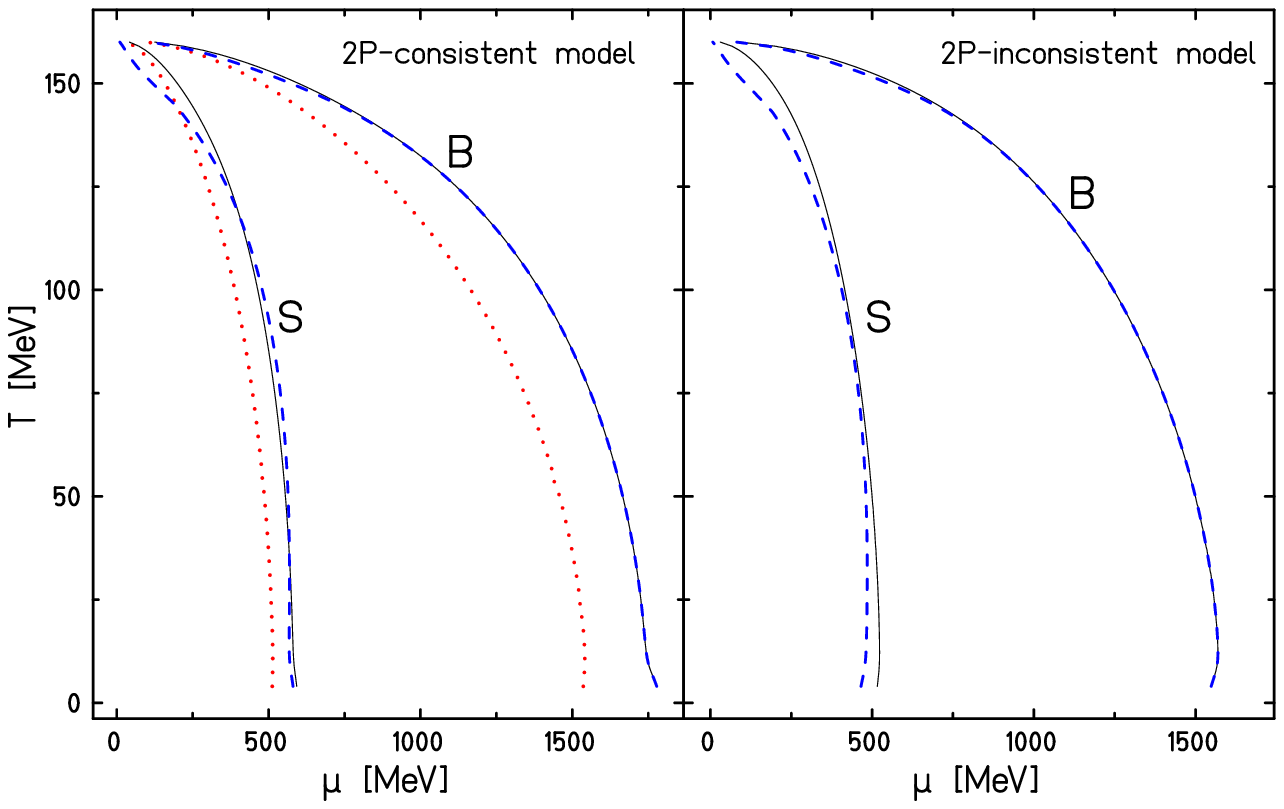}
\caption[C2]{ The phase diagram in the $T$--$\mu_B$ (marked by B)
and in the $T$--$\mu_S$ (marked by S) plane for the 2PC and 2PIN
models. The plasma and hadron boundaries are shown by full and
dashed--lines, respectively. The dotted--lines are the approximate
results obtained with $\mu_S=0$ and with $\mu_S$
   from Eq.(\ref{eqt14}).}
    \label{fig2}
\includegraphics[height=9.5cm, angle=-90,clip]{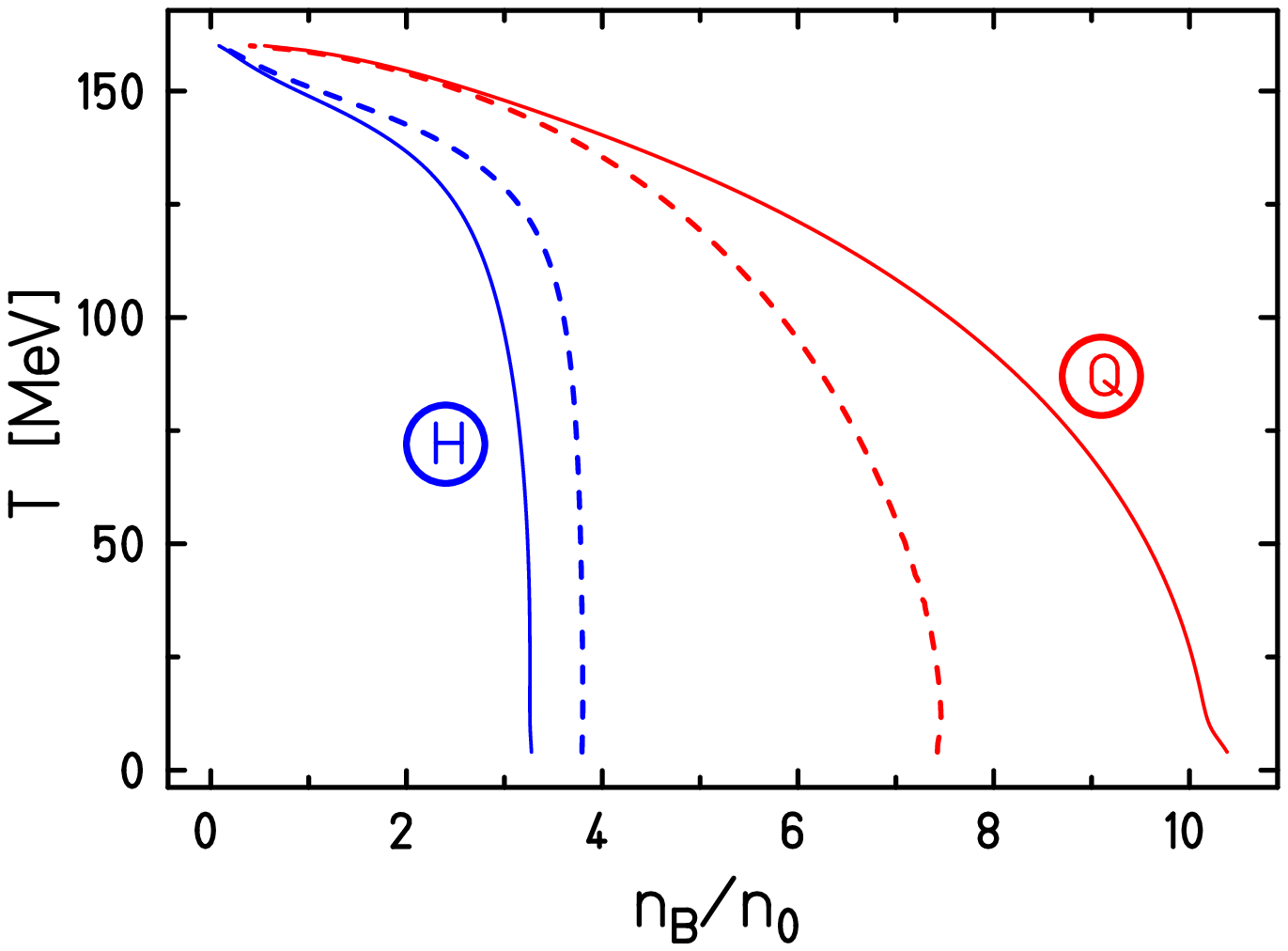}
\caption[C3]{ The phase diagram in the  ($T$--$n_B$) plane for
  2PC (full--lines) and 2PIN (dashed--lines) models. }
\label{fig3}
\end{center}
\end{figure}

The resulting phase boundaries in the $T$--$\mu$ plane are shown
in Fig.~\ref{fig2}. The difference in $\mu_B$ at the phase
boundaries
 described by Eqs.(\ref{eqt12a}--\ref{eqt12c}) is  small while for
the strange chemical potential $\mu_S$ it is more noticeable. It
is natural to expect that in the high temperature plasma
$\mu_S\approx\mu_B/3$. On the other hand, in the hadronic phase
and at low temperatures, where strangeness is carried mostly by
  kaons and $\Lambda$--hyperons, the strange chemical potential is
roughly approximated by $\mu_S\approx 0.5 \ (\mu_B +
m_K-m_{\Lambda})\approx 550\mbox{MeV}$. Both these expectations
are in agreement with our numerical results. Nevertheless, also in
the high temperature hadronic phase the strange chemical potential
exhibits an approximately linear dependence on the baryon chemical
potential.

In Figs.~\ref{fig2} and \ref{fig3}, the resulting phase diagrams
are shown in the $T$--$\mu_B$ and $T$--$\mu_S$ as well as
$T$--$n_B$ planes. The role of thermodynamical consistency is
particularly evident in the $T$--$n_B$ plane. As seen in
Fig.~\ref{fig3}, the baryon density $n_B$ at the plasma boundary
is increased while it is slightly decreased at the hadron side in
the 2PC model as compared with the 2PIN approach. Consequently,
the range of the coexistence region grows from $\sim (4 \div
7.5)n_0$ to $\sim (3.5 \div 10)n_0$.
\begin{figure}[htb]
\begin{center}
\includegraphics[height=9cm,clip]{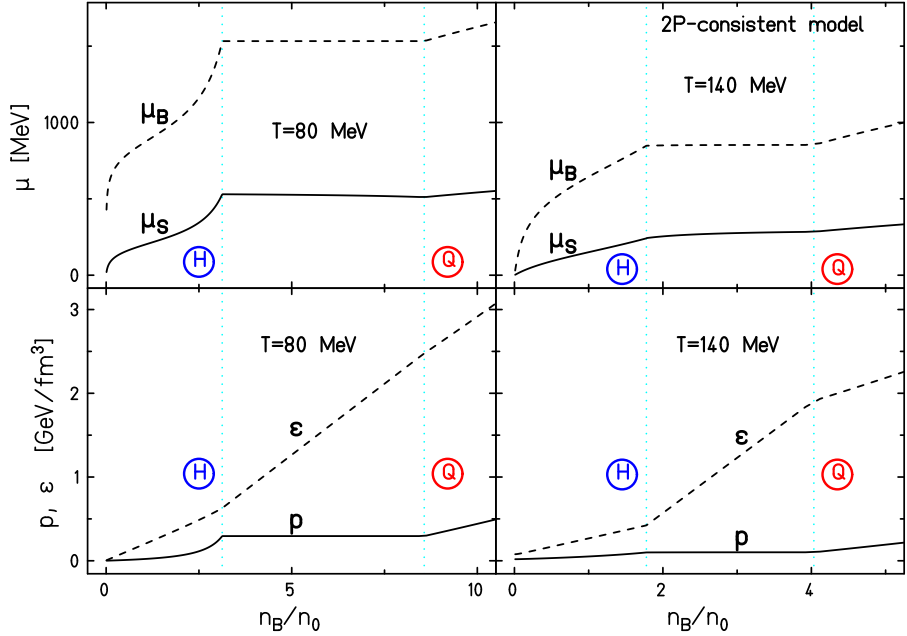}
\caption[C4]{ Dependence of different thermodynamical quantities
on the baryon density within  2PC model. The results are shown for
two different temperatures.  The  hadron and plasma phase
boundaries are shown by dotted--lines.} \label{fig4}
\end{center}
\end{figure}

\begin{figure}[htb]
\begin{center}
\includegraphics[height=9cm,clip]{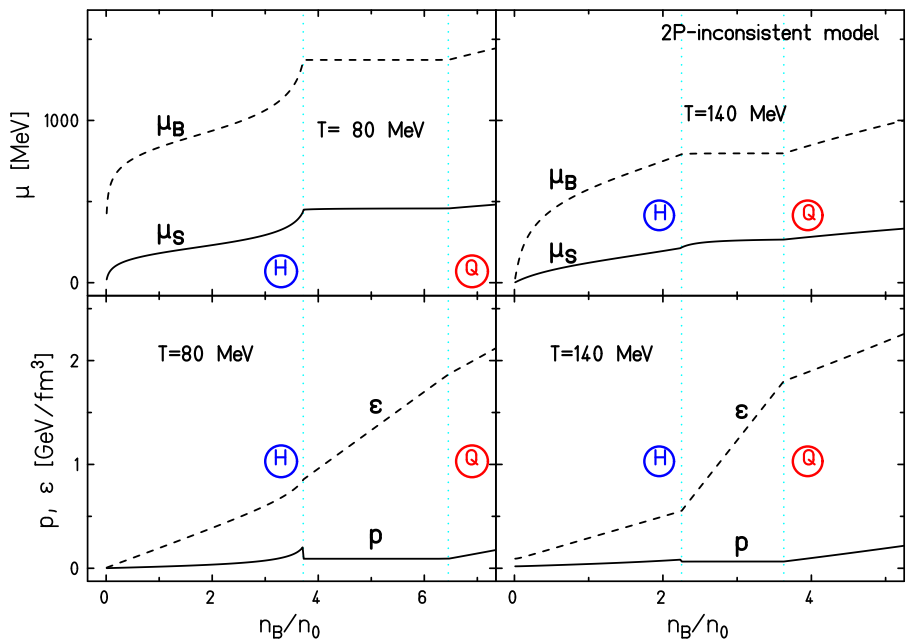}
\caption[C5]{The same as in Fig.~\ref{fig4} but for the
  2PIN model.}
\label{fig5}
\end{center}
\end{figure}

Thermodynamical properties and the differences  between the
two--phase bag models are shown in Figs.~\ref{fig4} and
\ref{fig5}. In both cases the baryon and strange chemical
potentials are continuous  when crossing the  phase boundaries.
This guarantees that   the system is chemically stable. Demanding
the conservation of strangeness   in each phase separately
\cite{LRBH86} would results in a discontinuity in $\mu_S$ . In
contrast to the case with only one conserved charge, the chemical
potentials  are not necessary constant within the Gibbs
coexistence region. Depending on the values  of $\mu$ at the
hadronic and plasma boundaries (see Fig.~\ref{fig2}), the chemical
potentials (in particular $\mu_S$) can be either increasing or
decreasing functions of $n_B$. Although this change is not large,
it influences the strangeness separation in the phase coexistence
region.

The energy density is seen in Figs.~\ref{fig4} and \ref{fig5} to
be a monotonously increasing function of $n_B$ in both models. The
pressure is also continuous within 2PC model and is  higher than
in the 2PIN approach. In addition, in the latter  model the
pressure also suffers a jump at the boundary of the hadronic
phase, which increases with decreasing temperature. Such an EoS
would lead to a mechanical instability of the hydrodynamic flow.
As seen in Figs.~\ref{fig4} and \ref{fig5}, the changes in
pressure across the coexistence region are quite small.
Consequently, the system expands very slowly. This is a specific
feature expected for the systems with a first order phase
transition.

We stress that there are at least two problems, which show up when
the EoS discussed above is employed in hydrodynamic calculations.
First, as shown in~\cite{RGSG91,PSS01}  causality is violated at
densities $n_B\gsim 3.5n_0$. Second, the ideal gas model with an
excluded volume correction does not reproduce the saturation
properties of nuclear matter. An attempt to combine  the excluded
volume correction with a mean field treatment of the hadronic
interactions resulted in an incompressibility parameter which is
too large $K \geq 550 \ MeV$~\cite{RGSG91}.

\subsection{Statistical mixed phase model}

The mixed phase (MP) model ~\cite{NST98,TNS98,NST96} is a
phenomenological model of the EoS with  a deconfinement phase
transition of QCD which   shows  a satisfactory agreement  with
the lattice data.
 The underlying assumption of the MP model is that
unbound quarks and gluons {\it may coexist} with hadrons forming a
{\it spatially homogeneous} quark/gluon--hadron phase which we
call  a generalized Gibbs mixed phase. Since the mean distance
between hadrons and quarks/gluons in the  mixed phase may be of
the same order as that between hadrons, the interactions between
all these constituents
 (unbound quarks/gluons and hadrons) play an important
role.  The strength of this interactions   defines the order of
the phase transition.

To find the free energy within the MP model \cite{NST98,TNS98},
the following effective Hamiltonian, expressed in terms of
quasiparticles interacting with a density--dependent mean field,
is used~:
\begin{eqnarray}
 H = \sum_i \sum_{\sigma} \int d{\bf r} \ \psi^+_i({\bf r},\sigma ) \
\left( \ \sqrt{-\nabla^2 + m^2_i}+U_i(\rho) \ \right) \
\psi_i({\bf r},\sigma ) - C(\rho ) V \ . \label{eqH}
\end{eqnarray}
Here $ \psi_i({\bf r},\sigma )$ denotes a field operator for the
quasiparticle species   $ i$ characterized by the  mass $m_i$ (the
current masses for quarks and gluons and the free hadron masses
are used here). The  index $\sigma$ accounts for spin, isospin and
color degrees of freedom. Furthermore,  $U_i$ is the mean field
acting on particles of type $i$,  $C(\rho)$ is a potential energy
term, which is needed to avoid double counting of the interaction,
and $V$ is the volume of the system.

By requiring  thermodynamical consistency
\cite{NST98,TNS98,GY95,BST01} one finds  constraints on the
parameters in the  Hamiltonian. The constraints follow
from~\cite{NST98,TNS98}
\begin{equation}
\langle \frac{\partial H}{\partial T} \rangle\, = \,0\, , \quad
\langle \frac{\partial H}{\partial \rho_i} \rangle\, = \,0 \;\; ,
\label{eq2}\end{equation} where $\langle \dots \rangle$ denotes
the statistical average.
 For the Hamiltonian (\ref{eqH}), these  conditions reduce to
\begin{eqnarray}
\sum\limits_i\rho_i
      \frac{\partial U_i}{\partial \rho_j}\; - \;
           \frac{\partial C}{\partial \rho_j}
               \;=\;0 \;\;,\quad
\sum\limits_i\;\rho_i
             \frac{\partial U_i}{\partial T}\;-\; \frac{\partial
  C}{\partial T}\;=\;0\,,
\label{eq3}
\end{eqnarray}
which, as shown in~\cite{NST98,TNS98} , imply that $U_i(\rho)$ and
$C(\rho )$ do not explicitly depend on temperature.

We model color confinement  by assuming the following density
dependence for the mean--field potential of quarks and gluons
\begin{equation}
U_q(\rho)=U_g(\rho)={A\over\rho^{\gamma}}~;
 \ \ \ \gamma >0
\label{eq6}
   \end{equation}
where
\begin{equation}
\rho=\rho_q + \rho_g +\sum_j \rho_j = \rho_q + \rho_g +\sum_j
\nu_j \ n_j \label{eq_rho}
\end{equation}
is the total number density of quarks and gluons in the local rest
frame and  $\rho_q$ and  $\rho_g$ are the number densities of
unbound (deconfined) quarks and gluons ($\rho_{pl}\equiv \rho_q +
\rho_g$), while $n_{j}$ is the number density of hadrons of type
$j$ having $\nu_j$  number of valence quarks inside. The presence
of the total number density  $\rho$ in (\ref{eq6}) implies
interactions between all components of  the generalized Gibbs
mixed phase. The  potential (\ref{eq6}) exhibits two important
limits of QCD. For $\rho \to 0$, the interaction potential
approaches infinity, {\em i.e.}  an infinite amount of energy is
necessary to create an isolated quark or gluon. This obviously
simulates confinement of colored objects. In the opposite limit of
large energy density,  $\rho \to \infty$, we have  $U_g\to 0$
which is consistent with asymptotic  freedom.

In the description of the hadron components, the  MP model
accounts not only for hadron--hadron  but also for
quark/gluon--hadron interactions. The mean field acting on the
hadron species $j$ in the MP model has two terms~:
\begin{eqnarray}
U_{j}=U_{j}^{(h)}+U_{j}^{(pl)}\;. \label{eq13}
\end{eqnarray}
In the limit where there are no unbounded quarks and gluons,
$U_{j}^{(pl)}=0$, i.e.,   $U_{j}=U_j^{(h)}$. This happens at low
densities, where colored degrees of freedom are confined in
hadrons.

Due to the constraints (\ref{eq3}) the second term in
Eq.(\ref{eq13}) my  be written as ~\cite{NST98}:
\begin{eqnarray}
U_{j}^{(pl)}\;=\;\frac{\nu_j\,A}{\rho^{\gamma}}
 \left(1-(1-w_{pl})^{-\gamma}\right)\;,
\label{eq17}\end{eqnarray} where $\displaystyle
w_{pl}=\rho_{pl}/\rho$  is the fraction of quark--gluon plasma in
the mixed phase~\footnote{We note that the resulting hadron single
particle potential becomes very attractive near the plasma phase
boundary, and even diverges in the plasma phase. We do not believe
that this behaviour is physical. Nevertheless, we adopt this
convenient prescription, since it does not have a noticeable
effect on the thermodynamics nor on the hydrodynamic flow.}. Thus,
if $U_q$ and $U_g$ are known, the thermodynamic consistency
conditions (\ref{eq3}) allow us to unambiguously determine  the
correction term $C(\rho$) in Eq. (\ref{eqH}).

The hadronic potential $U_{j}^{(h)}$ is described by a non-linear
mean-field mo\-del~\cite{Zim}
\begin{eqnarray}
U_{j}^{(h)}\;=  g_{r,j}\;\varphi_1 (x) + g_{a,j}\;\varphi_2 (y)\;,
\label{eqZ}
\end{eqnarray}
where $g_{r,j} > 0$ and $g_{a,j} < 0$ are repulsive and attractive
coupling constants, respectively.

Thermodynamic consistency implies that the functions
$\varphi_1(x)$ and $\varphi_2(y)$   depend only on particle
densities. In Ref. \cite{Zim} these functions are chosen such that
\begin{eqnarray} b_1
\varphi_1 = x, \quad -b_1 (\varphi_2 + b_2 \varphi_2^3 ) = y
\label{eq22}
\end{eqnarray}
where
$$ x=\sum\limits_{\nu_i} g_{r,i}\; \rho_{\nu_i},\quad
y=\sum\limits_{\nu_i} g_{a,i}\;\rho_{\nu_i}\;. $$ and  $b_1$ and
$b_2$ are  free parameters. In \cite{Zim} considering a mixture of
nucleons and $\Delta$'s the model parameters were fixed such that
to reproduce the saturation properties of nuclear matter and the
ratio of the $\Delta$ to nucleon coupling constants. We generalize
this approach by including all hadrons in our model and assuming
that the coupling constants scale with the number of constituent
quarks~:
\begin{eqnarray}
U_{j}^{(h)} = \nu_j\,\Bigl( \widetilde\varphi_1(\rho-\rho_{pl}) +
\widetilde\varphi_2(\rho-\rho_{pl})\,\Bigr)\;, \label{eq23}
\end{eqnarray}
where $\widetilde\varphi_1$ and $\widetilde\varphi_2$ satisfy the
equations
\begin{eqnarray}
c_1 \widetilde\varphi_1 = \rho-\rho_{pl}, \quad -c_2
\widetilde\varphi_2 - c_3 \widetilde\varphi_2^3 = \rho-\rho_{pl}\;
\label{eq24}
\end{eqnarray}
with $\rho-\rho_{pl}=\sum\nolimits_{\nu_j}\nu_j \rho_{j}$. The
parameters in Eq. (\ref{eq24})  are given by \cite{NST98}
$$ c_1
= \frac{b_1}{(g_{r,j}/\nu_j)^2}, \quad c_2 =
\frac{b_1}{(g_{a,j}/\nu_j)^2}, \quad c_3 = \frac{b_1
b_2}{(g_{a,j}/\nu_j)^4}\; $$ and are fixed by  requiring that the
properties of the ground state ($T=0$,
$n_B=n_0\approx0.17\;fm^{-3})$ of nuclear matter are reproduced:
i.e.  a binding energy per nucleon of $-16\; MeV$,
incompressibility of $210\;MeV$ and vanishing pressure.

We also  addressed to the extension of the Zimanyi
model~\cite{Zim} as { interacting hadron gas} (InHG) model with no
phase transition. The $\mu_B$--dependence of the pressure in this
model is illustrated in Fig.~\ref{fig1}.

The thermodynamics in the MP model is obtained from the partition
function  in the standard way.   The baryon and strange chemical
potentials are fixed by   the baryon number and strangeness
conservation,
\begin{eqnarray}
\label{eq18a}
n_B(T,\mu )&=& \sum_{j\in q,h}  b_j \ n_j(T,\mu_j )~, \\
n_S(T,\mu )=0 &=& \sum_{j\in q,h}  s_j \ n_j(T,\mu_j )~,
\label{eq18b}
\end{eqnarray}
where the sum is taken over all quarks, gluons and hadrons.
 The same set of  hadrons and resonances is  used here as in the previous models.

As an example we quote an expression for particle number density,
\begin{eqnarray}
n_j(T,\mu_j )= \frac{ g_j}{2\pi^2}\int_0^{\infty} dk k^2
 \left[   \exp \left( \frac{\sqrt{k^2 + m_j^2}- b_j\mu_B
-s_j\mu_S +U_j}{T} \right)
 \pm 1 \right]^{-1}.
\label{eq19}
\end{eqnarray}
In the energy density and pressure there are further terms,
originating from the interactions. These terms, are obtained
\cite{NST98,TNS98} by solving the consistency conditions
(\ref{eq3}).

\subsection{Thermodynamics in the mixed phase model}

The parameters of the MP model are fixed \cite{NST98,TNS98} by
requiring that the lattice results for thermodynamical quantities
in the pure gauge and two flavor QCD are reproduced. In this study
the quark--gluon interaction parameters $\gamma =0.62$ and
$\displaystyle A^{1/(3\gamma+1)} = 250$ MeV~ were obtained. The
same parameters are assumed to be valid also for the interactions
with strange quarks.

\begin{figure}[h]
\begin{center}
\includegraphics[height=9cm,clip]{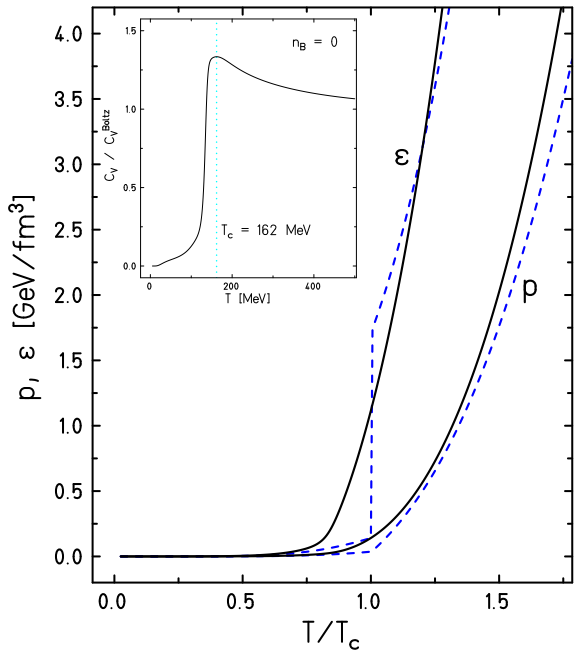}
\caption[C6]{Temperature dependence of the energy density and
  pressure at vanishing  total baryon density. Full and dashed--lines
  are the MP and 2PC model results, respectively. The insert figure
  shows  the reduced heat capacity.}
\label{fig6}
\end{center}
\end{figure}

In Fig.~\ref{fig6} we compare  the thermodynamical properties of
the MP and 2PC models. The energy density and pressure in the 2PC
model shows the typical behavior for a system with a first order
phase transition: an abrupt change in the energy density at
$T=T_c$ and smooth change in the pressure. In the MP model on the
other hand, both $\varepsilon$ and $p$ vary continuously with
temperature. The transition temperature in the MP model,
$T_c\approx 160$ MeV, is defined by the maximum of the heat
capacity (see insert in Fig. \ref{fig6}).

\begin{figure}[htb]
\begin{center}
\includegraphics[width=8cm, angle=-90,clip]{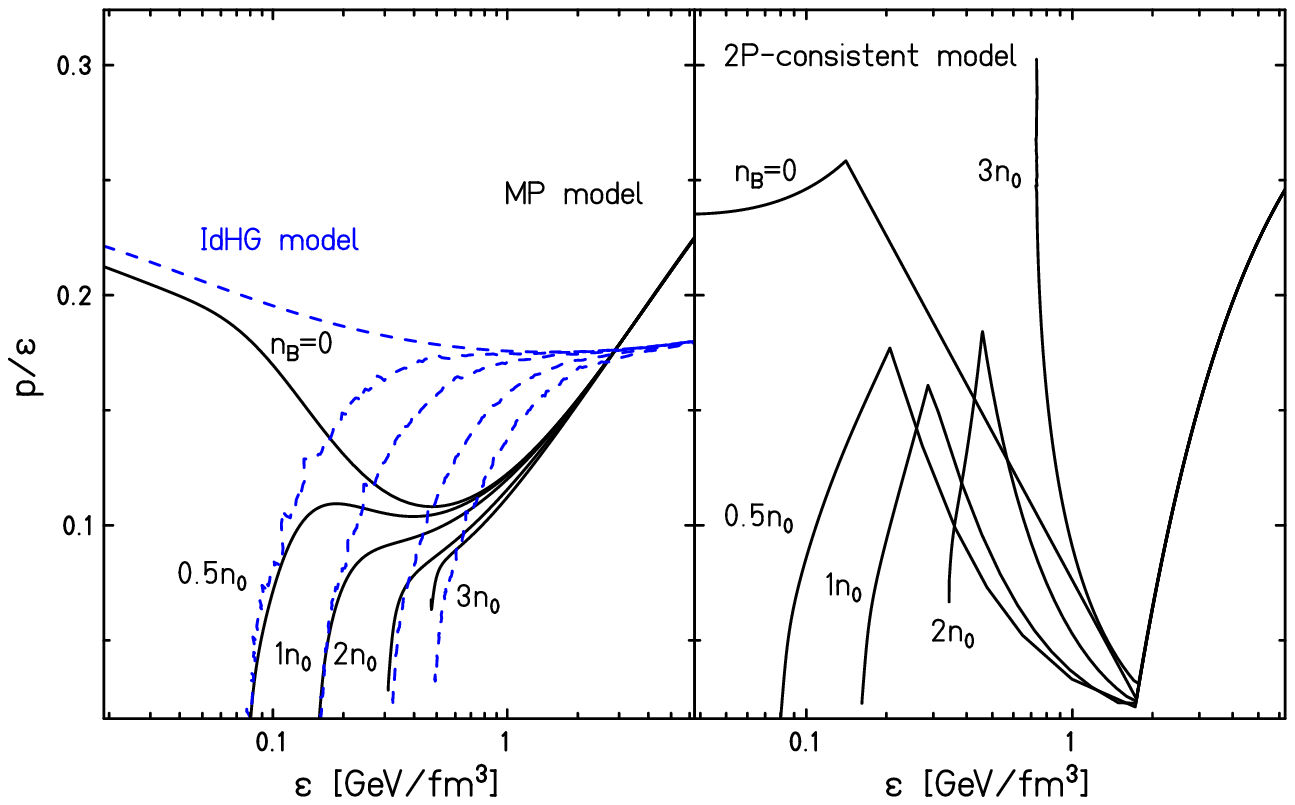}
\caption[C7]{The ratio of the pressure ($p$) to the energy density
($\epsilon$) as a function of $\epsilon$. The results are for
different values of a total baryon density ($n_B$) and for three
models of the EoS. } \label{fig7}
\end{center}
\end{figure}

 In Fig.~\ref{fig7} the ratio $p/\varepsilon$ is shown for different
values of the total baryon number density in three models under
consideration.
A common feature of these models is that for finite baryon
densities they all exhibit a clear threshold behavior. The
threshold is shifted to higher $\varepsilon$ with increasing
$n_B$. However, in contrast to the ideal hadron gas, both the MP
model and 2P model have the {\em softest point} in the EoS, i.e.,
a minimum \cite{HS95} of the function
$p(\varepsilon)/\varepsilon$. A particular feature of the MP model
is that even for $n_B=0$ the softest point is not very pronounced
and located at a relatively low energy density: $\varepsilon_{SP}
\approx 0.45$ GeV/fm$^3$. This is consistent with lattice result
~\cite{RS86}.
In the MP model, the softest point is gradually washed out with
increasing baryon density and vanishes completely for  $n_B \gsim
0.5 \ n_0$. This is, however, not the case in the 2P models, where
one finds a pronounced softest point at large energy density
$\varepsilon_{SP} \approx 1.5$ GeV/fm$^3$, which depends only
weakly on the baryon density $n_B$ see Fig.~\ref{fig7}).
Finally, in the  InHG model as well as  in the relativistic ideal
hadron gas  there is obviously no  softest point in the EoS.

The differences in the thermodynamical properties of the above
models will be also reflected in the expansion dynamics of a
thermal fireball created in heavy ion collisions. The effect of
these differences on strangeness production and evolution will be
explored in the following sections.

\section{Strangeness production}

\subsection{Strangeness content in equilibrium}
The conservation of strangeness in the coexistence region of
quarks and hadrons implies that the total number of strange and
antistrange quarks are equal. However, the $s$--$\bar s$  content
in the individual phases may differ from zero. The strangeness
content of the quarks in the mixed or plasma phase is
characterized  by two ratios~: $\rho_s/\rho_{\bar s}$ and
$D_s=(\rho_s+\rho_{\bar s})/\rho_{pl}$ (see Eq. (\ref{eq_rho})).
The second ratio gives the strangeness fraction in the plasma.

\begin{figure}[htb]
\begin{center}
\includegraphics[height=8.5cm,clip]{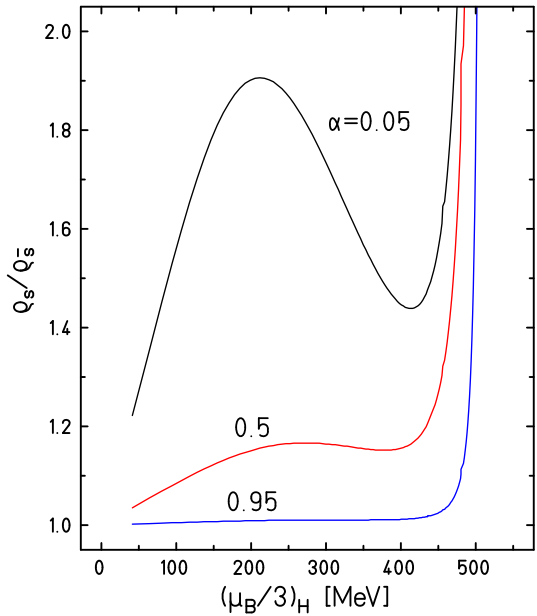}
\caption[C8]{ Ratio of strange to antistrange quark densities  in
a quark--gluon plasma   calculated along the hadronic boundary.
The results are for PC model calculated with different values of
the volume  fraction ($\alpha$) occupied by a quark--gluon plasma.
} \label{fig8}
\end{center}
\end{figure}

In Fig.~\ref{fig8} the ratio $\rho_s/\rho_{\bar s}$ is shown as a
function of $\mu_B$ for a fixed plasma fraction $\alpha$. For
$\alpha \sim 1$   the ratio $\rho_s/\rho_{\bar s}\approx 1$ for
almost all values of $\mu_B$. However, if  $\alpha << 1$, that is
when  the volume in the mixed phase is mostly occupied by hadrons,
the separation of strange and antistrange  quarks  is clearly seen
in Fig.~\ref{fig8}. This is mainly because the hadronic component
of the mixed phase is dominated  by the kaons, while the hyperons
are suppressed due to their large masses. This strangeness excess
through kaons is compensated by creation of $s$--quarks in the
plasma. The results in Fig.~ \ref{fig8} are in qualitative
agreement with Ref.~\cite{GKS87} where the 2PIN model without
higher mass resonances  was employed. The contribution  of higher
mass  resonances  results in an increase of $\rho_{s}/\rho_{\bar
s}$ for $(\mu_B/3)_H \approx 400-500 \ MeV$.

\begin{figure}[htb]
\begin{center}
\includegraphics[width=11.cm,clip]{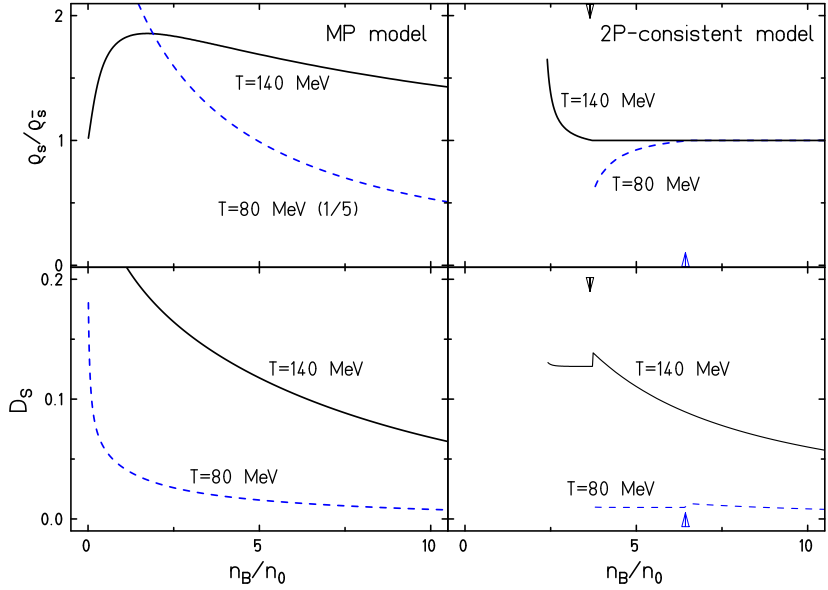}
\caption[C9]{The $\rho_s/\rho_{\bar s}$ ratio for quark component
and  strangeness  fraction ($D_s$)
  for unbound quarks as a function of baryon density.
The results are shown for different temperatures and for two EoS.
    The plasma boundary is marked  by arrows.  Note the factor 1/5 in the MP
  model at
  $T=80$ MeV.}
\label{fig9}
\end{center}
\end{figure}

In Fig. \ref{fig9} the strangeness composition in an equilibrium
system is compared for two different models.  In the bag model EoS
and at high temperature ($T\sim 140 \ MeV$) the $\rho_s/\rho_{\bar
s}$ ratio decreases when the baryon density inside the Gibbs mixed
phase approaches  the plasma boundary.
However, for the moderate temperatures ($T\sim 80 \ MeV$), the
ratio $\rho_s/\rho_{\bar s}<1$  and it increases with $n_B$. The
above behavior is a direct  implication  of the simultaneous
conservation of strangeness and the  baryon number. If these
conservation laws are decoupled \cite{GKS87}, then this behavior
at low temperatures is not seen.

In the MP model the  $\rho_s/\rho_{\bar s}>1$ for all values of
the baryon density. For a fixed temperature the $\rho_s/\rho_{\bar
s}$ ratio is  seen in Fig. \ref{fig9} to gradually decrease
 with increasing density. Its values are  noticeably higher than in
the 2P model.  In both models, however,  the strangeness
separation effect is stronger when the system is closer to the
hadronic boundary,   i.e. where there is   small admixture of
quarks. For the  2P
 model this  corresponds to the existence of a small blob of plasma
 while in the MP model
  a homogeneous admixture of unbound quarks and gluons with
 small concentration.

 Above the hadronic phase  boundary, the $n_B$--dependence of $D_s$
in the 2PC model is similar to that in  the MP model. The
strangeness fraction in the MP model is the largest below the
hadronic boundary and maximal in baryon free matter.
In Fig. \ref{fig9} we note a jump in $D_S$ which corresponds to a
jump in strange particle multiplicity when  crossing the phase
boundary; a similar jump is observed in the baryon number.

From the above discussion, it is clear, that the strangeness
content and its distribution in the transition region from the
quark--gluon plasma to the hadronic phase is strongly model
dependent. It is effected by the order of the phase transition and
the strength and the form of the interactions between
constituents. These differences are particularly evident at
moderate values of the temperature and baryon density. This is
just the region which is traversed by an expanding  system created
in  heavy ion collisions on its way towards the chemical
freeze--out. Thus, one could  expect that the order of the phase
transition and particular strangeness dynamics could  manifest
itself in observables in heavy ion collisions.

\subsection{Strangeness evolution in expansion dynamics}

To study the possible influence of the EoS on observables in heavy
ion collisions, we have to describe the space--time evolution of a
thermal  medium that is created in the initial state. This is
conveniently done within a hydrodynamical model. The EoS is an
input for constructing the energy---momentum tensor, which is
needed in the hydrodynamical equations.

To solve the  hydrodynamical equations for a given experimental
set--up one needs  to specify the initial conditions.   The initial
volume, the entropy and the baryon number densities in the
collisions  are modelled within QGSM transport code~\cite{QGSM}.
The predictions of this model are consistent with the results
obtained within the RQMD and UrQMD transport codes.

We assume that, in the center of mass frame, the initial state is
a cylinder of  radius $R=5 \ fm$ and Lorentz contracted length
$L=2R/\gamma_{c.m.}$.  This initial state  corresponds to the time
when the centers of the colliding nuclei just have passed the
point of full overlap.\footnote{ A detailed  description of the
procedure to fix the  initial conditions in heavy ion collisions
can be found in~\cite{NST98,TNS98,GY95,BST01}.} We neglect the
transverse expansion and assume that the hydrodynamical evolution
of the fireball is described by a one-dimensional isentropic
expansion of the scaling type in the longitudinal direction. In
this approximation the entropy and baryon density decrease
inversely proportional to the expansion time. The values of all
other thermodynamic quantities are  obtained from the EoS at each
temporal step (see, for example \cite{SSW86}).

\begin{figure}[htb]
\begin{center}
\includegraphics[width=8.cm, angle=-90]{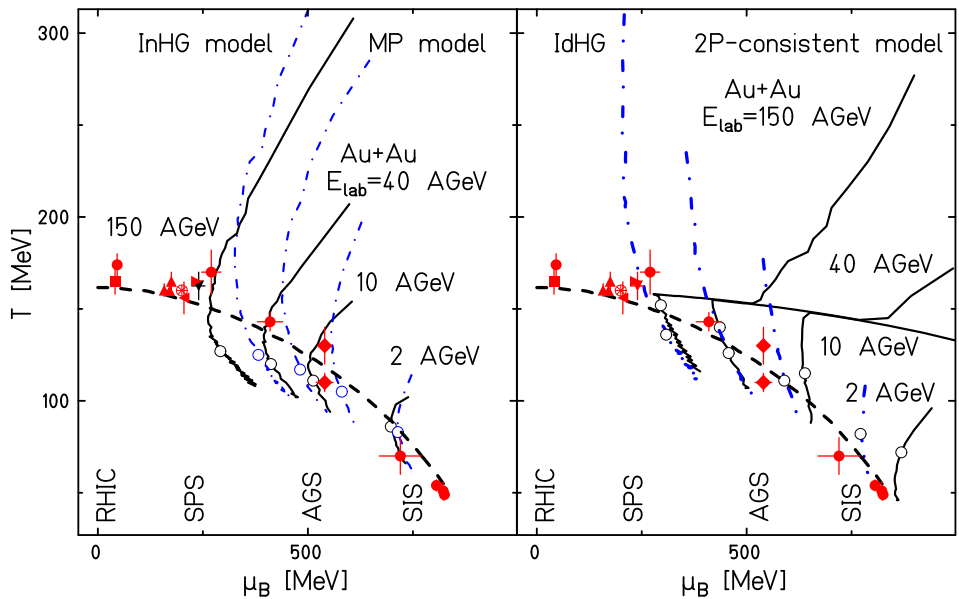}
\caption[C10]{ A compilation of the  chemical freeze--out
parameters  from Refs. \cite{Red01,CR98,Stock02} obtained with the
hadron resonance gas partition function at different
beam--energies (filled dots, squares and triangles). The smooth
dashed curve is the universal freeze--out curve of fixed
$\left<E_{had}\right>/\left<N_{had}\right> \simeq 1 \ GeV$ from
Ref. \protect\cite{CR98}. Also shown are dynamical trajectories
for central $Au$--$Au$ collisions calculated within different
models (the interacting hadron gas model (InHG), the mixed-phase
model (MP), the ideal hadron gas model (IdHG) and the
thermodynamically consistent two--phase model (2PC)). The empty
circles near the end of each trajectory correspond to freeze--out
condition of fixed energy density, $\varepsilon_f\simeq 0.135 \
GeV/fm^3$.
 } \label{fig10}
\end{center}
\end{figure}

In Fig. \ref{fig10} we show the fireball evolution trajectories
for central $Au$--$Au$ collisions in the $T$--$\mu_B$ plane for
different collision energies and for different EoS. The chemical
freeze--out parameters obtained \cite{rev,Red01,CR98,Stock02}
within the statistical model at  different collision energies are
also shown in this figure. Clearly, the chemical freeze--out
parameters from SIS up to RHIC are well described by the universal
condition of fixed energy/particle,
$\left<E_{had}\right>/\left<N_{had}\right>\simeq 1$ GeV
\cite{CR98,CR99}.

The dynamical trajectories show a strong dependence on the
properties of the EoS. In the MP model there is a turning point
seen in all trajectories,  { i.e.} the point where $\partial T
/\partial \mu_B$ changes sign. The existence of such a point is a
general feature of the MP  model and  is directly related to the
appearance of two limiting regimes:

(i) At high temperatures and in the ultra--relativistic limit,
$m_q\to 0$, the thermodynamic potential $\Omega=-Vp$ can be
obtained analytically from Eqs.(\ref{eqt5}) and (\ref{eqt8})

\begin{equation}
\Omega= -V \ (a_1T^4+a_2T^2\mu_B^2+a_3\mu_B^4)~.
\label{eq20}
\end{equation}
The entropy per baryon
\begin{eqnarray}
\frac{s}{n_B}=\frac{\partial {\Omega}/\partial T}{ \partial{\Omega}/
  \partial \mu_B} = \frac{2a_1 +a_2 \ (\displaystyle
  \frac{\mu_B}{T})^2}{a_2 \ (\displaystyle \frac{\mu_B}{T}) +a_3 \
  (\displaystyle \frac{\mu_B}{T})^3}\,,
\label{eq21}
\end{eqnarray}
is conserved along trajectories defined by $\mu_B/T={\it const}$.
Thus, in the high temperature limit, an isentropic expansion is
characterized by a linear relation between $T$ and $\mu_B$.

(ii) At intermediate temperatures, the system can be approximated
by a Boltzmann gas (\ref{eqtB}) of a non--relativistic nucleons.
In this case the entropy in the dilute gas approximation is given
by
\begin{eqnarray}
S&=&-\frac{g_N V}{(2\pi )^3} \int d^3p \ \left[ f \ln f +(1-f)\ln (1-f) \right]
  \nonumber \\ &\approx &N_B \left[ 1- \frac{ \int d^3p \ f \ln f }{ \int
      d^3p \ f  } \right]
\label{eq21a}
\end{eqnarray}
with the  distribution function $f=\exp [(\mu_B -m_B
-p^2/2m_B)/T]$. In this temperature range, conservation of the entropy per baryon
implies that
\begin{equation}
\frac{s}{n_B}= \frac{5}{2} +\frac{m_B-\mu_B}{T}= {\it const}~.
\label{eq25}
\end{equation}
Thus, for intermediate temperatures,  we again find a linear
relation between $T$ and $\mu_B$ but with a negative slope. The
different behavior of $\mu_B(T)$ at high and intermediate
temperatures, implies that there is a turning point in the fireball
expansion trajectories, as seen in Fig. \ref{fig10}.

The dynamical trajectories calculated in the MP model pass quite
close to the phenomenological freeze--out points. For all
collision energies, the turning point is located on the universal
freeze--out curve of fixed energy/particle. This fact has been
noticed already in Ref. \cite{TCN01} for the MP model with two
light quarks. The contribution of strange quarks and the
requirement of strangeness conservation modifies the dynamical
expansion path of the fireball. This is particularly evident for $
E_{lab}\lsim 10 $ AGeV where neglecting the strange quarks gives
rise to a visible shift of the turning point towards smaller
$\mu_B$.

In the parameter range below the phenomenological freeze--out curve
the expansion path in the MP, IdHG as well as in MP model are quite
similar. In the InHG model, however, there is a small shift toward
larger values of $\mu_B$. This agreement indicates that in the
final stage the expansion path depends only weakly on details of
the equation of state. The dynamical path is, to a large extent,
determined by the entropy/baryon and strangeness conservation which
in the hadronic phase puts strong constraints on the particle
composition of the fireball. In this case the space time evolution
and thermodynamics is governed by a gas of weakly interacting
resonances, the effective degrees of freedom in the low temperature
phase of QCD. This may be the reason behind the success of the
non--interacting hadron resonance gas in the description of bulk
observables in heavy ion collisions.

The differences between the various equations of state in the
evolution of the thermal fireball is clearly visible above the
freeze--out curve. In contrast to the MP model, the IdHG turning
points do not correlate with the freeze--out curve. There is also
no softest point in the InHG and IdHG model. The dynamical
trajectories within the 2P bag--models  exhibit a characteristic
re--heating regime in the phase coexistence region. For this model,
the expansion trajectory closely follows the phase boundary in this
regime, as shown in Fig. \ref{fig2}. At SPS energies and above, the
hadronic end of the intermediate coexistence region in the
$T$--$\mu_B$ plane (the so--called "hottest hadronic point") is
close to the phenomenological chemical freeze--out point. At lower
energies there is no such correlation for the 2P models. For
$E_{lab}\lsim 10 \ AGeV$ the initial state is in the phase
coexistence region.

\begin{figure}[htb]
\begin{center}
\includegraphics[height=9.cm,clip]{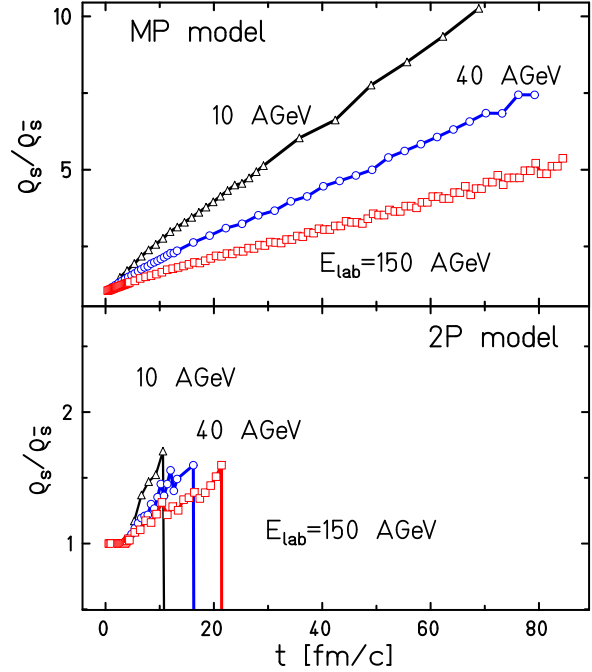}
\caption[C11]{  Time evolution of the ratio of strange to
anti--strange quark densities  in the hadronic
  component. The   MP and 2P model results are
for central  $Au$--$Au$ collisions at different beam--energies.}
\label{fig11}
\end{center}
\end{figure}

The question of strangeness separation in heavy ion collisions
addressed  in Ref.~\cite{GKS87} for static system  can be
reanalyzed in our approach for  dynamically evolving fireball. The
results are shown in Fig. \ref{fig11} for Au--Au collisions at
different bombarding energies within 2P and MP models. In both
models $\rho_{s}/\rho_{\bar s} >1$, since there is no chance for
the system to pass  through a high density baryonic state where
$\rho_s$ could be less then unity.

In the 2P model we find that stran\-ge\-ness is separated to a
less degree at the exit point from the phase coexistence region
than found  in~\cite{GKS87}. On the other hand in the MP model the
system evolves much longer and consequently  a higher degree of
strangeness separation is obtained. This effect is stronger at
$E_{lab}=10 \ AGeV$  than at $160 \ AGeV$.

So far the differences between various models for the expansion
dynamics were discussed on the level of global thermodynamical
quantities. It is of particular interest to explore physical
observables that are directly measured in heavy ion collisions. In
the following we consider strange particle multiplicity ratios to
discuss the influence of the equation of state on particle yields.
The predictions of the different models will be compared at {\it
thermal freeze--out} where the particle momentum distributions are
frozen. We assume a shock--like freeze--out~\cite{Bug96} where
energy, the total baryonic  and strangeness charges are conserved.
The thermal freeze--out conditions are assumed to be determined by
the fixed energy density $\varepsilon_f  \approx 0.9 n_0 m_N =0.135
\ GeV/fm^3$. Below this energy density the system consists of a free
streaming gas of non interacting particles. The  thermal
freeze--out points are shown in Fig. \ref{fig10} by empty circles
on each trajectory for all models and for all collision energies.

\begin{figure}[tbh]
\begin{center}
\includegraphics[width=8.cm, angle=-90,clip]{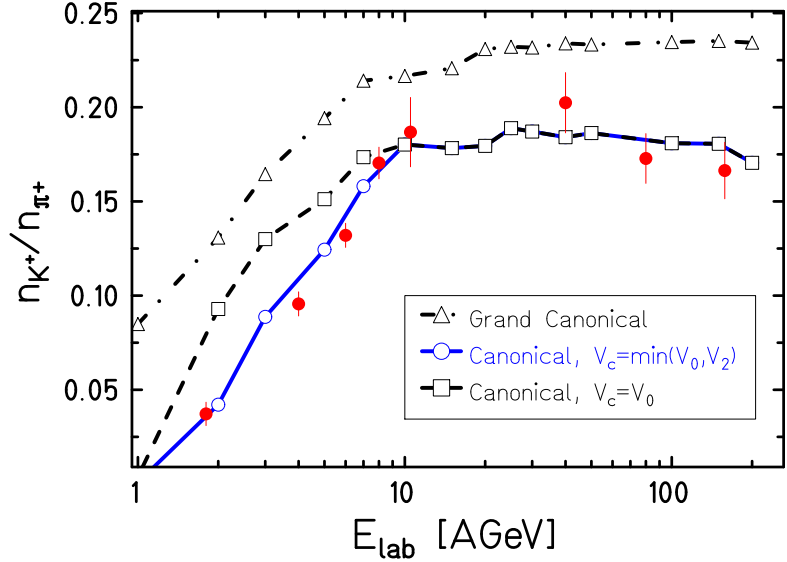}
\vspace{-3mm} \caption[C12]{  The $K^+/\pi^+$ ratio
 as a function of a beam--energy.  The data points are from
 ~\protect\cite{Stock02}.
The lines are the MP model results obtained  in the grand
canonical as well as canonical formulation of strangeness
conservation for different parameterizations of the
  volume parameter $V_c$ (see text). }
\label{fig12}
\end{center}
\end{figure}

The excitation function of the relative yields of  $K^+$ mesons
calculated within the  MP model is  shown in  Fig. \ref{fig12} as
triangles. For reference, we also show in this figure the
$4\pi$--integrated data for  $K^+/\pi^+$ ratio   obtained in heavy
ion collisions at different beam energies. The shape of the kaon
excitation function  in the MP model is similar to that seen in
the data. However, the absolute values  are overestimated,
especially for  the low collision energies. We have to stress,
however, that the models discussed here are still not quite
suitable to be compared with data. First, the conservation of
electric charge  was not taken into account. The isospin asymmetry
is particularly relevant at low collision energies (below AGS)
where it can change the charge particle multiplicity ratios by up
to 20$\%$. Second, the hydrodynamical model applied here describes
a longitudinally expanding fireball. This is, to a large extend
sufficient at RHIC energies, however, it may be not valid at AGS
or SIS where transverse expansion cannot be neglected.
Furthermore, only part of particle mass spectrum was included with
the  masses up to 1.6 GeV.  At AGS and higher energies the
contributions from heavier resonances increases the yields of
lighter particles. Finally, the system may be out of chemical
equilibrium at some stages during the evolution from chemical
towards thermal equilibrium \cite{shuryak}. Nevertheless, all
these effects cannot account for the observed discrepancy by a
factor of five between the MP model and data at low collision
energies (Fig. \ref{fig12}). However, the differences may be due
to the grand canonical (GC) treatment of the strangeness
conservation used in the calculations.

\noindent

In the GC ensemble strangeness is conserved on the average and is
controlled by the strange chemical potential. Within  the
statistical approach, the use of the grand canonical ensemble for
particle production can be justified only if the number of
produced particles that carry a conserved charge is sufficiently
large. In this case also event--averaged multiplicities can be
treated in a grand canonical formulation. In this approach, the
net value of a given charge (e.g. electric charge, baryon number,
strangeness, charm, etc.) fluctuates from event to event. These
fluctuations can be neglected (relative to the mean particle
multiplicity) only if the particles carrying the charges in
question are abundant. Here, the charge is indeed conserved on the
average and a grand canonical treatment is adequate. However, in
the opposite limit of low production yields (as is the case for
strangeness production in low energy heavy-ion collisions) the
particle number fluctuation can be of the same order as the event
averaged value. In this case  charge conservation has to be
implemented exactly in each event \cite{rev,r20}. In the
statistical physics the exact conservation of quantum numbers
requires a canonical (C) formulation of the partition function.

The grand canonical  $Z^{GC}$ and canonical $Z^C_S$ partition
functions are connected by a cluster decomposition in the fugacity
parameter ($\lambda \equiv \exp (\mu_s/T) $)
\begin{equation}
 Z^{GC}(T,V,\mu_B,\lambda)=
 \sum_{s=-\infty}^{s=\infty} \lambda^s\  Z_s^{C}(T,V,\mu_B)~.
\label{eq26a}
\end{equation}
The relation (\ref{eq26a}) can be inverted and the canonical
partition function with  total strangeness  $S=0$ is obtained from
\begin{equation}
Z_{S=0}^{C}(T,V,\mu_B)=\frac{1}{2\pi} \int_{0}^{2\pi}d\phi \
 Z^{GC}(T,V,\mu_B,\lambda \to e^{i\phi})~.
\label{eq26b}
\end{equation}
Neglecting the contributions from multistrange hyperons and
assuming Boltzmann statistics the density of kaons in the C
ensemble is given by \cite{HR85,CRS91}
\begin{equation}
n_K^C  = n_K^B \ \frac{{\cal S}_1}{\sqrt{{\cal S}_1{\cal S}_{-1}}}
\ \frac{I_1(x)}{I_0(x)}~. \label{eq30}
\end{equation}
where the argument of the Bessel function $I_s(x)$ is
\begin{equation}
x\equiv 2\sqrt {{\cal S}_1{\cal S}_{-1}}. \label{eq29}
\end{equation}
with
$${\cal S}_s=V_c\sum_j n^B_j~. $$
Here  the particle density $n^B_j$  for hadron species  $j$ is
given by Eq. (\ref{eqtB}) with $\mu_j=\mu_B b_j$. The  sum is
taken over all particles and resonances carrying strangeness $s$.
The volume  $V_c$ is a model parameter which is interpreted  as
the strangeness correlation volume.\footnote{For a more detailed
discussion of the interpretation and the role of this parameter

see e.g. Ref. \cite{rev}.} In the equilibrium statistical model a
correlation volume $V \equiv V_1 \simeq 1.9\pi A_{part}/2$ was
found to reproduce the experimental multiplicity ratios for all
measured particle. In our dynamical approach, $V_c$ is assumed to
be the initial volume of the collision fireball
$V_c=V_0(E_{lab})$, and thus is energy dependent \cite{ed}.

From  Eqs.(\ref{eq30}) and (\ref{eqtB}) it is clear that grand
canonical and  canonical results for the kaon density are related
by the substitution  \cite{COR99}:
\begin{equation}
 \exp (\mu_s/T) \to  \frac{{\cal S}_1}{\sqrt{{\cal S}_1 {\cal
       S}_{-1}}} \ \frac{I_1(x)}{I_0(x)}~.
\label{eq31}
\end{equation}
Thus, the main difference between C and GC results is contained in
a reduction of the fugacity parameter by the factor $F\equiv
I_1(x)/I_0(x)$. In the limit of a large volume $V_c$, i.e., $x\to
\infty$, the ratio $ F\to 1$ and the GC and C results coincide.
However, in the opposite limit, $x\to 0$, the factor  $F < 1$
which leads to a suppression of the strange particle densities.
The canonical suppression depends strongly on the temperature and
the correlation volume. Both these parameters, in particular the
temperature,  are dependent on the collision energy.

\begin{figure}[htb]
\begin{center}
\includegraphics[height=10.cm, angle=-90,clip]{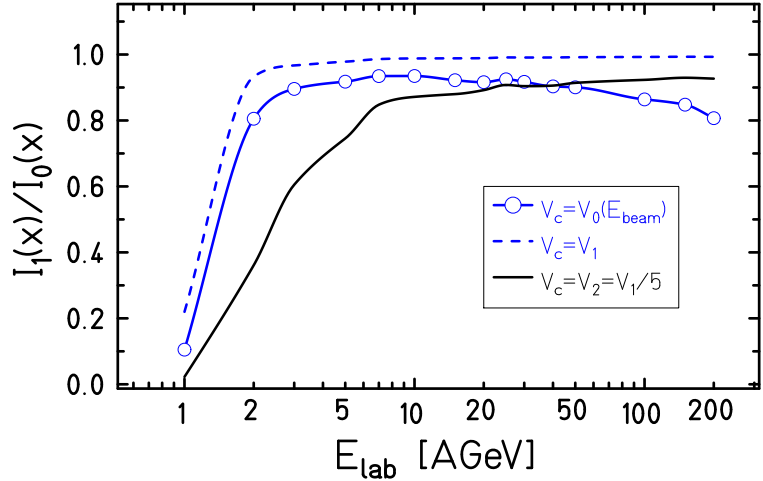}
\caption[C13]{The  beam--energy dependence of the strangeness
suppression factor  for central $Au$--$Au$ collisions. This factor
is calculated within MP model at freeze--out for  different
parameterizations of the correlation volume $V_c$ (see  text).
       }
\label{fig13}
\end{center}
\end{figure}

The sensitivity of the suppression factor on $E_{lab}$ and the
correlation volume is shown in Fig. \ref{fig13}. The canonical
suppression factor calculated with $V_c=V_1$  is increasing with
collision energy and reaches  its asymptotic value at $E_{lab}
>10$ GeV. Obviously the magnitude of the suppression is
strongly dependent on the correlation volume. This is shown in
Fig. \ref{fig13} for three different parameterizations of $V_c$.
Particularly interesting is the behavior of $F$ for
$V_c=V_0(E_{lab})$. Due to the Lorentz contraction the initial
volume is decreasing with increasing collision energy. At lower
collision energies this decrease of the  volume is compensated by
an increase of temperature such that the suppression factor
increases with $E_{lab}$. However for $E_{lab}>10$ GeV there is
only a moderate increase of freeze--out temperature that is not
sufficient to overcome a decrease of $V_0$. Consequently for
$E_{lab}>10$ GeV the suppression factor starts to decrease with
energy.

The amount of canonical suppression at fixed $E_{lab}$ also
depends strongly on the temperature which in turn is determined by
the energy density at freeze--out. In the equilibrium analysis of
particle production at SIS \cite{COR99} the energy density at
chemical and thermal freeze--out was a factor of three lower than
the value used in the present dynamical study, $\epsilon \simeq
0.135$ GeV/fm$^3$. Consequently, for $V_c=V_0$ and for
$1<E_{lab}<2$ GeV the canonical suppression found in Ref.
\cite{COR99} was much stronger than that shown in Fig.
\ref{fig13}. We have not tuned the parameters to reproduce
previous results. In low energy heavy ion collisions  the
expansion trajectories and the freeze--out parameters will change
once the transverse expansion is taken into account.

In fig. \ref{fig12} we show the effect of the canonical
suppression on the $K^+$ excitation function  calculated in the MP
model with two different parameterizations of the correlation
volume: $V_c=V_0$ and $V_c={\rm min} (V_0,V_2)$ where $V_2=V_1/5$.
As expected, there is a noticeable decrease of $K^+$ yield  due to
the exact treatment of strangeness conservation. The suppression
of strangeness at energies beyond AGS is entirely due to the
energy--dependent Lorentz contraction of the initial correlation
volume. In Fig. \ref{fig12} the results of a calculation also are
presented where the choice of $V_c={\rm min} (V_0,V_2)$ is
optimized to reproduce the $K^+/\pi^+$ data.

\begin{figure}[h]
\begin{center}
\includegraphics[width=10.3cm,clip]{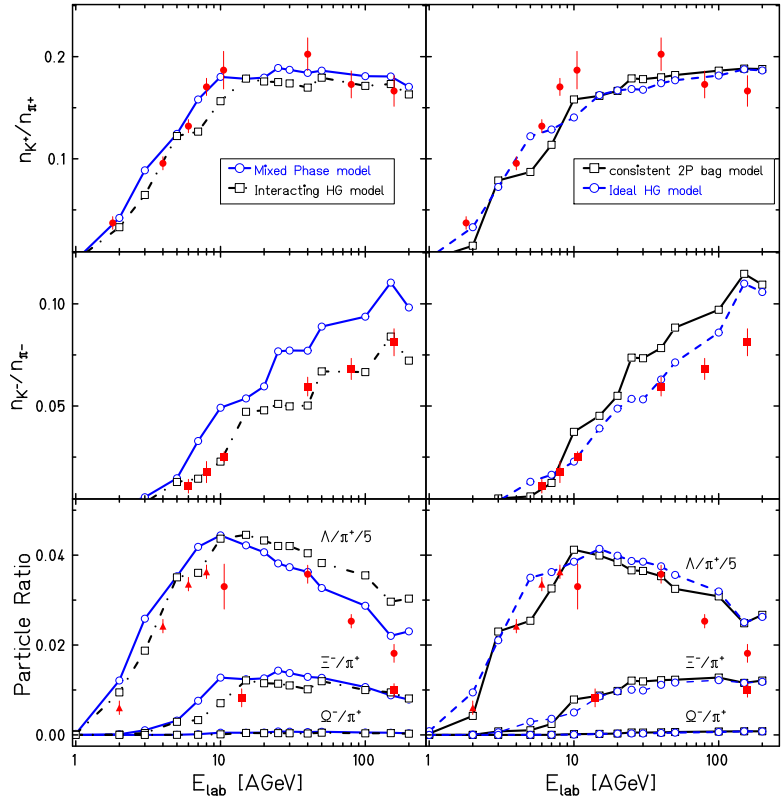}
\vspace{-3mm} \caption[C14]{The ratios of $4\pi$--integrated
strange particle yields per pion yields   for central $Au$--$Au$
collision as a function
 of  beam--energy. The  compilation of  experimental
 data is taken from~\cite{Stock02,Redlich01}.
 The calculated excitation functions are for  different  EoS with
 the canonical suppression factor. }
\label{fig14}
\end{center}
\end{figure}

The above analysis of $K^+$ excitation function clearly shows that
due to associated strangeness production and the small production
cross sections at low collision energies one has to implement
exact strangeness conservation. In the following we will implement
this concept in all models and discuss the predictions for
strangeness production and energy dependence. In the calculations
we use the correlation volume $V_c={\rm min} (V_0,V_2)$.

In Fig. \ref{fig14} we calculate relative excitation functions for
different strange mesons and baryons for four hydrodynamical
models. The most striking result seen in this figure is that all
models yield very similar results for the strangeness  excitation
functions. This is particularly true for the production of
$K^+/\pi^+$ and $\Lambda/\pi^+$ where the results of all models
besides InHG, are hardly distinguishable. Some differences are seen
on the level of $K^-$ excitation function which are mainly due to
larger sensitivity of $K^-/\pi^-$ ratio to the value of the
temperature. It is interesting to note that all models show a
maximum in the $\Lambda /\pi$ excitation function for
$10<E_{{lab}}<30$ GeV. Such a maximum is found also in equilibrium
models \cite{BMCOR02}.

\begin{figure}[htb]
\begin{center}
\includegraphics[height=10.cm, angle=-90,clip]{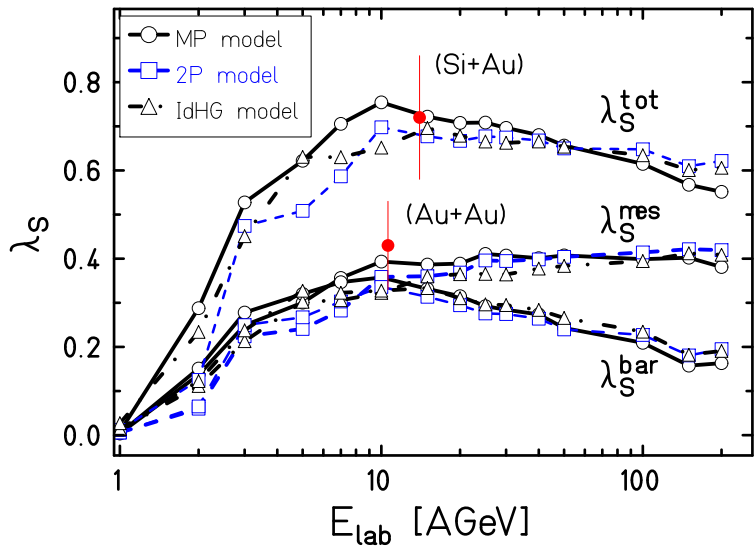}
\caption[C15]{  The Wroblewski  ratio $\lambda_s$ as a function of
    beam--energy  for central $Au$--$Au$ collisions. The contributions of
   mesons and baryons are shown separately. The  points at AGS energies are
   from ~\cite{Stock02}.}
\label{fig15}
\end{center}
\end{figure}

The relative strangeness content of the produced particles in heavy
ion collisions is  characterized by the Wr\`oblewski factor
\cite{BMCOR02,wrob},

\begin{equation}
\lambda_S=\frac{2 <s\bar{s}>}{<u\bar{u}>+<d\bar{d}>} \label{eqWr}
\end{equation}
where the quantities in angular brackets refer to the number of
newly created quark--antiquark {\it pairs}. The  Wroblewski factor
is shown   in Fig. \ref{fig15}  for different collision energies.
The separate contributions to $\lambda_S$ from  strange mesons and
baryons as well as its overall value is calculated within the MP,
2P and IdHG models. The results are  compared with  $\lambda_S$
obtained in an equilibrium model analysis of experimental data at
AGS energies. There is a  surprising agreement of all dynamical
models on the relative strangeness content of the fireball at
freeze--out. The results are also consistent with  the equilibrium
model  \cite{BMCOR02}. However, the maximum spread of the
Wroblewski factor seen in Fig. \ref{fig15} is broader  than
previously seen in the equilibrium canonical model  \cite{BMCOR02}.
In the dynamical models  there is also a small shift in the
position of this maximum towards lower energy.

\section{Summary and conclusions}

The main objective of this article was  to explore the influence
of the expansion dynamics, the equation of state and the nature of
deconfinement phase transition on strangeness production  in heavy
ion collisions.

We have discussed and formulated different models for a phase
transition in hight density QCD matter. The thermodynamical
properties of these models and  the role of the order of the phase
transition as well as the interactions between the particles has
been analyzed. We have addressed the question of the Gibbs
construction of the phase transition in the presence of two
conserved charges and emphasized the problem of causality and
thermodynamical consistency.

The strangeness separation in the transition region from the
quark--gluon plasma to the hadronic phase was also studied. The
asymmetry in the relative concentration of strange and
anti--strange quarks  in the hadronic and quark--gluon component
in the phase coexistence region was found in all models that
exhibit a phase transition. However, the largest effect  was
observed  in the mixed--phase model with  a crossover--type
deconfinement phase transition.

The differences in equilibrium thermodynamics of the  models were
studied on the dynamical level. We have shown that the
hydrodynamical expansion  trajectories of the fireball in the
$T$--$\mu_B$ plane are very sensitive to the equation of state. We
considered the effect of the different expansion paths on
strangeness production. Our detailed analysis show that there is
almost no sensitivity of strangeness observables on the equation of
state or on the expansion trajectories. This was demonstrated for
several strange particle excitation functions.

To relate the model predictions with experimental data   we have
extended our study to a canonical formulation of strangeness
conservation. We have discussed the phenomenological limitations
of our dynamical models and the possible extension needed to
provide a quantitative description of the observed  particle
yields in heavy ion collisions.

Exact strangeness conservation substantially reduces the strange
particle yields in heavy ion collisions for $E_{lab}<10$ GeV. For
higher energies a moderate  suppression is also found if the
beam--energy dependence of the volume parameter $V_c$ is taken
into account. We have shown that the assumption that $V_c$ is the
volume of the initially produced Lorentz contracted fireball may
lead to a negative slope in the energy dependence of the
$K^+/\pi^+$ ratio. However, within considered models,  the almost
singular behavior of the excitation function near $E_{lab}<20$ GeV
for $K^+/\pi^+$ ratio found recently by the NA49
collaboration~\cite{Gazd03} was not reproduced. Simplified
hydrodynamics with the assumption of a shock--like particles
freeze--out in heavy-ion collisions results in a very smooth
behavior of the strange particle excitation functions.

  \vspace*{0.5cm}

{\section*{ Acknowledgements}

Stimulating  discussions with  Yu.~Ivanov  are gratefully
acknowledged. We also thank J.~Knoll, E.~Kolomeitsev, A.~Parvan,
A.~Shanenko and D.~Vo\-skre\-senski for useful  comments. E.G.N.
and V.D.T. acknowledge the hospitality at the Theory Group of GSI,
where  this work has been done. This work was supported in part by
DFG (project 436 RUS 113/558/0-2) and RFBR (grant 03-02-04008).
K.R. acknowledges the support of the Alexander von Humboldt
Foundation (AvH) and the Polish State Committee for Scientific
Research (KBN) grant   2P03 (06925).

 \end{document}